\documentclass[%
 aip,jcp,
 amsmath,amssymb,
 reprint,
]{revtex4-2}

\usepackage[a-1b]{pdfx} 

\usepackage{graphicx}% Include figure files
\usepackage{dcolumn}% Align table columns on decimal point
\usepackage{bm}% bold math
\usepackage[utf8]{inputenc}
\usepackage[T1]{fontenc}
\usepackage{mathptmx}
\usepackage{listings}
\usepackage{rotating} % Rotating table

\definecolor{RED}{HTML}{FF0000}
\usepackage{color}
\usepackage{dcolumn} % decimal align in tables
\usepackage{bm} % bold math
\usepackage{graphicx}
\usepackage{multirow} % for table cells to span rows
\usepackage{pifont} % for checkmarks
\usepackage{epsfig}
\usepackage{amsmath} % matrix
\usepackage{subfigure}
\usepackage{float}
\usepackage{booktabs}
\usepackage{tabularx}
\usepackage{natbib}
\usepackage{gensymb}
\usepackage{rotating}
\usepackage{threeparttable}
\usepackage{comment}
\usepackage[normalem]{ulem}

\newcommand{\mb}{\mathbf}

\begin{document}

\author{A. Eugene DePrince III}
\email{adeprince@fsu.edu}
\affiliation{
             Department of Chemistry and Biochemistry,
             Florida State University,
             Tallahassee, FL 32306-4390 USA}

\author{Stephen H. Yuwono}
\affiliation{
             Department of Chemistry and Biochemistry,
             Florida State University,
             Tallahassee, FL 32306-4390 USA}   
             
\author{Henk Eshuis}
\affiliation{
Department of Chemistry and Biochemistry,
Montclair State University, 
Montclair, New Jersey 07043, USA}

\title{Cavity Quantum Electrodynamics Ring Coupled Cluster and the Random Phase Approximation}

\begin{abstract}
It is well known that the ground-state correlation energy from the particle-hole channel of the random phase approximation (RPA) is formally equivalent to that from a simplified coupled cluster doubles (CCD) model that includes only ring diagram contraction contributions in the residual equations [{\em J. Chem. Phys.} {\bf 129}, 231101 (2008)]. We generalize this analytic result to the cavity quantum electrodynamics (QED) case and demonstrate the numerical equivalence of QED-RPA and a QED ring-CCD model that accounts for double electron excitations, coupled single electron excitations / single photon creation, and double photon creation.
\end{abstract}

\maketitle

\section{Introduction}

The random phase approximation (RPA)\cite{Pines51_625,Bohm52_338,Pines53_609} and coupled-cluster (CC) theory\cite{Coester58_421,Kuemmel60_477,Cizek66_4256,Cizek69_35,Shavitt72_50,Li99_1,Musial07_291} share deep formal connections that underpin the systematic treatment of correlated many-body systems in nuclear physics, quantum chemistry, materials science, and condensed-matter physics. Numerical evidence of these connections was first put forward nearly 50 years ago,\cite{Freeman77_5512} whereas the formal proof demonstrating the equivalence of the ground-state correlation energy from a ring-diagram simplification of CC with double excitations (CCD) and the particle-hole channel of the RPA was developed only much more recently.\cite{Sorensen08_231101} Since the seminal work in Ref.~\citenum{Sorensen08_231101}, some of the same authors\cite{Bulik13_104113} and others\cite{Yang13_104112} have shown similar equivalencies exist between a particle-particle channel of the RPA and a ladder-diagram simplification of CCD. These connections extend to excitation energies,\cite{Berkelbach18_041103,Bartlett20_234101} where eigenvalue equations derived from the particle-hole or particle-particle RPA equations are equivalent to the equations of equation-of-motion (EOM) CC\cite{Emrich81_379,Bartlett89_57,Bartlett93_7029} models limited to the relevant excitation manifolds ({\em i.e.}, particle-hole or particle-particle and hole-hole).

{In the quantum chemistry community, correlated methods such as the RPA, CC, or EOMCC are usually applied to correlations among a single type of particle, {\em i.e.}, electrons. Over the last few years, though, a great deal of effort has gone into extending the standard electronic structure tools to consider both electron and photon degrees of freedom quantum mechanically, on equal footing.\cite{DePrince23_041301} Along these lines, cavity quantum electrodynamics (QED) generalizations of many familiar electronic structure methods have emerged, spanning single-reference many-body wave function approaches,\cite{Narang23_arXiv:2307.14822, Reichman24_1143,Koch20_041043, Manby20_023262, Corni21_6664, DePrince21_094112, Flick21_9100, Koch21_094113, Koch22_234103, Flick22_4995, Rubio22_094101, Knowles22_204119, DePrince22_054105, Rubio23_2766, Rubio23_10184, Koch23_4938, Koch23_8988, Koch23_031002, DePrince23_5264,DePrince24_064109,Koch24_8876,Stopkowicz24_9572,Koch24_8838,Koch24_e1684,Koch25_021040,Dreuw23_124128,Kowalski26_294} multireference theories,\cite{DePrince22_053710,Yu24_032804,Foley24_1214,Foley25_8812,Ronca25_6862,Huo23_2353,Huo24_16184,Foley24_174105} and, of course, density functional theory (DFT). \cite{Bauer11_042107,Rubio14_012508,Rubio15_093001,Rubio18_992,Appel19_225,Rubio19_2757,Narang20_094116,Rubio22_7817, Rubio23_11191,Narang23_383,DePrince22_9303,Rubio22_094101,DePrince23_5264, DePrince24_064109,Bauer11_042107,Rubio14_012508,Tokatly13_233001,Rubio17_3026,Tokatly18_235123,Varga22_194106,Shao21_064107, Shao22_124104, Yuwono25_8024} These developments come in response to a number of experimental studies demonstrating that strong light--matter coupling, which is facilitated by interactions between matter degrees of freedom and the quantized radiation fields associated with optical cavity modes, can lead to unexpected and non-trivial effects that can be leveraged in a variety of contexts, {\em e.g.}, to exert control over chemical transformations\cite{Ebbesen12_1592, Ebbesen16_11462, Shegai18_eaas9552, Ebbesen19_615, Ebbesen19_15324, George19_10635, Uji-i20_5332, Ebbesen20_10436, Borjesson21_2010737, Ebbesen21_16877, Shalabney21_chemrxiv.7234721.v5, George21_379, Ebbesen21_5712, George22_195, Jeffrey22_429} or modify materials properties.\cite{Whittaker98_6697,Mugnier04_036404,Bramati07_106401,Ebbesen15_1123,Ebbesen16_2403,Bellessa19_173902,Ebbesen16_7352,KenaCohen18_119,Singer19_1801682,Ebbesen21_1486,Singer21_085307,Giebink22_7937,Borjesson18_2273,Borjesson21_3255}}

{Among the {\em ab initio} cavity QED models that have been developed, the QED-CC approach\cite{Koch20_041043, Manby20_023262, Corni21_6664, DePrince21_094112, Flick21_9100, Koch21_094113, Koch22_234103, Flick22_4995, Rubio22_094101, Knowles22_204119, DePrince22_054105, Rubio23_2766, Rubio23_10184, Koch23_4938, Koch23_8988, Koch23_031002, DePrince23_5264,DePrince24_064109,Koch24_8876,Stopkowicz24_9572,Koch24_8838,Koch24_e1684,Kowalski26_294} stands out as robust and systematically improvable framework for simulating subtle effects stemming from cavity-molecule interactions. At the same time, as in standard electronic structure theory, the steep scaling of even low-order QED-CC models precludes their application to large molecules. As such, cheaper methods, such as QED-DFT, may offer the most realistic path toward {\em ab inito} simulations of large cavity-embedded systems. Unfortunately, several studies\cite{DePrince21_094112,DePrince23_5264,DePrince24_064109} have shown that QED-DFT models that ignore electron-photon correlation effects tend to overestimate cavity-induced changes to electronic structure, as compared to predictions made using wave function methods that explicitly account for such correlations ({\em e.g.}, QED-CC). While the development of accurate electron-photon correlation functionals is an active area of research,\cite{Rubio15_093001,Rubio17_113036,Rubio18_992,Rubio21_e2110464118,Tokatly23_235424,Rubio24_052823,Flick25_073002} it is surprising that RPA-based models have seemingly received no attention, even at a formal level, given the RPA's well-known role in standard electronic structure theory, \cite{Toulouse05_012510,Rubio06_073201,Kurth00_16430,Burke05_094116,Furche08_114105,Eshuis12_1084,Furche12_084105,Scheffler12_7447} where over the past few decades it has seen a revival as a ground-state correlation method. The RPA is non-perturbative and size extensive, it seamlessly includes long-range interactions and electronic screening, and it can be implemented at mean-field computational scaling.\cite{Furche08_114105,Furche12_084105,Scheffler12_7447} These features make the RPA a promising target for the description of cavity-induced changes to electronic structure in large systems.}

{Here}, we generalize the results of Ref.~\citenum{Sorensen08_231101} and demonstrate the equivalence of ground-state correlation energies obtained from a cavity QED version of the RPA and a ring-diagram simplification of a QED-CCD model that includes the effects of double electron excitations, coupled single electron excitations / single photon creation, and double photon creation. As such, this work places QED-RPA within the {QED-CC hierarchy and positions the method as a rigorous correlation model for electron-photon interactions in strongly-coupled systems. The remainder of this paper is organized as follows. Section~\ref{SEC:THEORY} provides the working equations of the QED-RPA and QED-ring-CCD approaches, followed by the proof connecting the them. Section~\ref{SEC:NUMERICAL_VALIDATION} then provides a numerical evidence of the equivalence of these methods, as well as an brief study of the impact of photon-photon correlation effects in these models. Some concluding remarks can be found in Sec.~\ref{SEC:CONCLUSIONS}.}

\section{Theory}
\label{SEC:THEORY}

\subsection{Cavity QED Hartree-Fock}

We begin with the Pauli-Fierz Hamiltonian,\cite{Spohn04_book,Rubio18_0118} which captures interactions between electronic degrees of freedom and quantized radiation fields. In the the length gauge and within the dipole approximation, this Hamiltonian takes the form
\begin{align}
        \label{EQN:PFH}
\hat{H}_\text{PF}
    &= \hat{H}_\text{e} + \omega_\text{cav} \hat{b}^\dagger\hat{b} - \sqrt{\frac{\omega_\text{cav}}{2}} {\bm \lambda}\cdot \hat{\bm{\mu}} \left(\hat{b}^\dagger + \hat{b}\right) \nonumber \\
    &+ \frac{1}{2} \left({\bm \lambda}\cdot \hat{\bm{\mu}}\right)^2
\end{align}
Here, the first and second terms represent the usual electronic Hamiltonian and the Hamiltonian for a single optical cavity mode, respectively. Note that this choice is made out of convenience and that the Hamiltonian is generalizable to the case of many cavity modes. The operators $\hat{b}$ and $\hat{b}^\dagger$ represent photon annihilation and creation operators, respectively, and $\omega_\text{cav}$ is the frequency of the cavity mode. The third term captures the coupling between the molecular degrees of freedom and the cavity mode, the strength of which is parametrized by the coupling vector, $\bm{\lambda}$. The symbol ${\bm \mu} = {\bm \mu}_\text{e} + {\bm \mu}_\text{nu}$ represents the molecular dipole operator, which includes both electronic and nuclear contributions. The last term is the molecular dipole self-energy. 

The mean-field approximation to the lowest-energy eigenstate of $\hat{H}_\text{PF}$ is the QED Hartree-Fock (HF) wave function
\begin{equation}
\label{EQN:QED_HF}
    |\Phi_{0}\rangle = |0_\text{e}\rangle \hat{U}_\text{CS}|0_{\rm p}\rangle
\end{equation}
where $|0_\text{e}\rangle$ is a Slater determinant of molecular electronic spin orbitals, $|0_{\rm p}\rangle$ is the photon vacuum, and $\hat{U}_\text{CS}$ is the coherent-state (CS) transformation operator. The photon part of the QED-HF wave function can be defined analytically with the choice\cite{Koch20_041043}
\begin{equation}
\label{EQN:U_CS}
    \hat{U}_{\rm CS} = {\rm exp}\left( \frac{-{\bm \lambda} \cdot \langle {\hat{\bm{\mu}}} \rangle }{\sqrt{2 \omega_{\rm cav}}} \left (\hat{b}^{\dagger} - \hat{b} \right ) \right) 
\end{equation} 
which diagonalizes the photon part of $\hat{H}_\text{PF}$ for any single electronic Slater determinant, $|0_\text{e}\rangle$. Here, the expectation value of the dipole operator is taken with respect to $|0_\text{e}\rangle$. For the determination of $|0_\text{e}\rangle$ and subsequent correlation treatment, it will be convenient to transform the Hamiltonian to the CS basis using this operator, giving
\begin{align}
\hat{H}_\text{CS} &= \hat{U}_\text{CS}^\dagger \hat{H}_\text{PF} \hat{U}_\text{CS} \nonumber \\
        \label{EQN:PFH_COHERENT}
     &= \hat{H}_\text{e} + \omega_\text{cav} \hat{b}^\dagger\hat{b} - \sqrt{\frac{\omega_\text{cav}}{2}} {\bm \lambda}\cdot [\hat{\bm{\mu}}_\text{e} - \langle \hat{\bm{\mu}}_\text{e}\rangle ] \left(\hat{b}^\dagger + \hat{b}\right) \nonumber \\
    &+ \frac{1}{2} \left({\bm \lambda}\cdot [\hat{\bm{\mu}}_\text{e} - \langle \hat{\bm{\mu}}_\text{e}\rangle ]\right)^2
\end{align}
Taking the expectation value of $\hat{H}_\text{CS}$ with respect to $| 0_\text{e}\rangle |0_\text{p}\rangle$ gives us the QED-HF energy
\begin{align}
    \label{EQN:QED_HF_ENERGY}
    E_\text{0} 
    = \langle 0_\text{e}| \hat{H}_\text{e} |0_\text{e}\rangle + \frac{1}{2} \langle 0_\text{e} | \left({\bm \lambda}\cdot[\hat{\bm{\mu}}_\text{e} - \langle \hat{\bm{\mu}}_\text{e}\rangle ]\right)^2 |0_\text{e}\rangle
\end{align}
which, we can see, depends only on the electronic part of the wave function. As such, $|0_\text{e}\rangle$ and $E_0$ can be determined from a modified HF procedure involving the isolated electronic Hamiltonian plus the DSE part of $\hat{H}_\text{CS}$.

\subsection{Cavity QED Random Phase Approximation}

\label{sec:qed_rpa}

Given $E_0$ and $|0_\text{e}\rangle$ determined via the QED-HF procedure, we can determine approximate excitation energies with a cavity QED generalization of the RPA. The QED-RPA eigenvalue problem takes the form\cite{Shao21_064107,DePrince22_9303}
\begin{widetext}
\begin{equation}
\label{EQN:CASIDA}
    \left ( \begin{array}{cccc}
    \mb{A} + \bm{\Delta} & \mb{B} + \bm{\Delta}' & \mb{g} & \mb{\tilde{g}} \\
    \mb{B} + \bm{\Delta}' & \mb{A} + \bm{\Delta} & \mb{\tilde{g}} & \mb{{g}} \\
    \mb{g}^\dagger & \mb{\tilde{g}}^\dagger &     \omega_\text{cav} & 0 \\
    \mb{\tilde{g}}^\dagger & \mb{{g}}^\dagger &    0  & \omega_\text{cav}
    \end{array} \right)
    \left( \begin{array}{c}
    \mb{X} \\
    \mb{Y} \\
    \mb{M} \\
    \mb{N} 
    \end{array} \right)
    =
    \left ( \begin{array}{cccc}
    \mb{1} & 0 & 0 & 0 \\
    0 & \mb{-1} & 0 & 0  \\
    0 & 0 & \mb{1} & 0   \\
    0 & 0 & 0 & \mb{-1} 
    \end{array} \right)
    \left( \begin{array}{c}
    \mb{X} \\
    \mb{Y} \\
    \mb{M} \\
    \mb{N} 
    \end{array} \right)
    \bm{\Omega},
\end{equation}
\end{widetext}
Here, $\bm{\Omega}$ is a diagonal matrix containing the positive-energy solutions to the QED-RPA problem, $\mb{X}$ and $\mb{Y}$ are electronic excitation and de-excitation amplitudes, $\mb{M}$ and $\mb{N}$ are photon creation and annihilation amplitudes, and $\mb{A}$ and $\mb{B}$ are the usual RPA matrices, with elements
\begin{align}
\label{eqn:rpa_a}
    A_{ai,bj} &= (\epsilon_a - \epsilon_i)\delta_{ij}\delta_{ab} + \langle ib ||aj\rangle \\
\label{eqn:rpa_b}
    B_{ai,bj} &= \langle ij || ab\rangle
\end{align}
The labels $i, j, ...$ / $a, b, ...$ refer to canonical molecular spin orbitals ($\phi$) that are occupied / unoccupied in the QED-HF reference function. The symbol $\epsilon_p$ refers to $p$th diagonal element of the Fock matrix, which has elements
\begin{align}
    F_{pq} = h_{pq} -( {\bm{\lambda}}\cdot \langle {\bm{\hat{\mu}}}_{\rm e}\rangle ) d_{pq} - \frac{1}{2} q_{pq} \nonumber \\
    +\sum_{i}\left ( \langle ip||iq\rangle + d_{ii}d_{pq} - d_{iq}d_{pi}\right ) 
\end{align}
where the indices $p$ and $q$ refer to general (occupied or virtual) molecular spin orbitals. The symbol $\langle pq||rs\rangle = \langle pq|rs\rangle - \langle pq|sr\rangle$ is an antisymmetrized electron repulsion integral (ERI) in physicists' notation, $h_{pq}$ refers to an element of the core Hamiltonian matrix, and $d_{pq}$ and $q_{pq}$ represent dressed electric dipole and quadrupole integrals that derive from the DSE part of the Hamiltonian and are defined by
\begin{align}
        d_{pq} &= - \sum_{\mu \in \{x,y,z\}} \lambda_\mu \int \phi^*_p r_\mu \phi_{q} d\tau \\
        q_{pq} &= - \sum_{\mu\nu \in \{x,y,z\}} \lambda_\mu \lambda_\nu \int \phi^*_p r_\mu r_\nu \phi_{q} d\tau
\end{align}
Here, $\lambda_\mu$ is a cartesian component of ${\bm{\lambda}}$, and $r_\mu$ is a cartesian component of the position vector.
The coupling vectors, $\mb{g}$, and counter-rotation coupling vectors, $\mb{\tilde{g}}$, have elements
\begin{equation}
\label{eqn:rpa_g}
g_{ai} = \tilde{g}_{ai} = -\sqrt{\frac{\omega_{\rm cav}}{2}} d_{ai} 
\end{equation}
Lastly, ${\bm \Delta}$ and ${\bm \Delta^\prime}$ are two-electron matrices arising from the DSE part of the Hamiltonian, with elements
\begin{eqnarray}
\label{EQN:DELTA}
\Delta_{ai,bj} = d_{ai} d_{jb} - d_{ab} d_{ij} \\
\label{EQN:DELTA_PRIME}
\Delta^\prime_{ai,bj} = d_{ai} d_{bj} - d_{aj} d_{ib}
\end{eqnarray}
{A derivation of the QED-RPA eigenvalue problem using Rowe's equation of motion formalism\cite{Rowe68_153} is provided in Ref.~\citenum{DePrince22_9303}.}

The correlation energy in standard RPA is obtained as a difference in excitation energies from the standard RPA problem and the Tamm-Dancoff approximation (TDA) to that problem (with $\mb{B}\to \mb{0}$). The QED-RPA correlation energy is obtained in a similar manner, except that two approximations are required to arrive at the second eigenvalue problem. We invoke both the TDA (with $\mb{B} + \bm{\Delta}^\prime\to \mb 0$) and the rotating wave approximation (RWA), in which we ignore the counter-rotation terms terms ($\mb{\tilde{g}} \to \mb 0$), leading to
\begin{equation}
\label{EQN:TDA}
    \left ( \begin{array}{cccc}
    \mb{A} + \Delta & \mb{{g}} \\
    \mb{g}^\dagger & \mb{\omega}_\text{cav} 
    \end{array} \right)
    \left( \begin{array}{c}
    \mb{\bar{X}} \\
    \mb{\bar{M}}
    \end{array} \right)
    =
    \left( \begin{array}{c}
    \mb{\bar{X}} \\
    \mb{\bar{M}}
    \end{array} \right)
    \bm{\bar{\Omega}},
\end{equation}
which is referred to as the TDA-RWA approximation in Ref.~\citenum{Shao21_064107}.
In analogy to standard RPA, the QED-RPA correlation energy is obtained from the difference in excitation energies from these two models:
\begin{align}
\label{eqn:qed_rpa_energy}
    E_\text{c}^\text{RPA} = \frac{1}{2}\left (\text{Tr}\left ( \bm{\Omega}\right) - \text{Tr}\left ( \bm{\bar{\Omega}}\right)\right) 
\end{align}
where we note again that the matrix $\bm{\Omega}$ only contains the positive excitation energies from the QED-RPA problem. 

\subsection{Cavity QED Ring Coupled Cluster Doubles}

\label{sec:qed_ring_ccd}

An alternative correlation treatment is provided by cavity QED-CC theory.\cite{Koch20_041043} We consider a QED-CC with doubles approach, where the wave function is
\begin{align}
|\Psi_\text{CC}\rangle = \text{exp}(\hat{T}) |0_\text{e}\rangle |0_\text{p}\rangle
\end{align}
with
\begin{align}
    \hat{T} = \frac{1}{4} \sum_{ijab} T^{2,0}_{ai, bj}\hat{a}^\dagger_a\hat{a}^\dagger_b\hat{a}_j\hat{a}_i + \sum_{ia} T^{1,1}_{ai} \hat{a}^\dagger_a \hat{a}_i \hat{b}^\dagger + \frac{1}{2}T^{0,2}\hat{b}^\dagger\hat{b}^\dagger
\end{align}
Here, $T^{2,0}_{ai,bj}$ is a cluster amplitude corresponding to double electronic transitions, $T^{1,1}_{ai}$ is an amplitude for single electronic transitions coupled to single photon creation, $T^{0,2}$ is a double photon creation amplitude, and $\hat{a}_p$ / $\hat{a}_p^\dagger$ are the usual fermionic annihilation / creation operators associated with $\phi_p$. For the full QED-CCD model, these amplitudes are determined by solving projective equations of the form
\begin{align}
\label{eqn:ccd_2}
    \langle \Phi_{ij}^{ab}|\langle 0_\text{p}|{\bar{H}}|0_\text{p}\rangle | 0_\text{e}\rangle &= 0 \\
\label{eqn:ccd_1}
    \langle \Phi_{i}^{a}|\langle 1_\text{p}|{\bar{H}}|0_\text{p}\rangle | 0_\text{e}\rangle &= 0 \\
\label{eqn:ccd_0}
    \langle 
    0_\text{e}|\langle 2_\text{p}|{\bar{H}}|0_\text{p}\rangle | 0_\text{e}\rangle &= 0
\end{align}
where $\bar{H} = e^{-\hat{T}} \hat{H}_\text{CS} e^{\hat{T}}$ is the similarity-transformed coherent-state basis Hamiltonian, $|n_\text{p}\rangle$ is a photon-number state containing $n$ photons, and $|\Phi_i^a\rangle$ and $|\Phi_{ij}^{ab}\rangle$ are Slater determinants of molecular electronic spin orbitals that are singly and doubly substituted relative to $|0_\text{e}\rangle$, respectively. Explicit, programmable expressions for Eqs.~\ref{eqn:ccd_2}-\ref{eqn:ccd_0} are provided in the Supporting Information. Here, we present equations for the ``ring'' simplification of QED-CCD (QED-rCCD), where we retain only the familiar ring-type electronic contractions, as well as electron-photon ring contractions corresponding to electron excitation / photon annihilation or photon creation / electron de-excitation. 

The ring simplification of the ${\bf T}^{2,0}$ residual equation is
\begin{widetext}
\begin{align}
    0 &= \langle ab||ij\rangle % B(ai,bj)
    + T_{ai,bj}^{2,0}(\epsilon_a + \epsilon_b - \epsilon_i - \epsilon_j) 
    + \sum_{ck} \langle ic||ak \rangle T_{ck,bj}^{2,0} 
    + \sum_{ck} \langle kb||cj\rangle T_{ai,ck}^{2,0} % T.A
        + \sum_{cdkl} T_{ai,ck}^{2,0} \langle kl||cd\rangle T_{dl,bj}^{2,0} \nonumber \\ % T.B.T
&+\Delta^\prime_{ai,bj}
    + \sum_{ck} \Delta_{ai,ck} T_{ck,bj}^{2,0}
    + \sum_{ck}   \Delta_{ck,bj} T_{ai,ck}^{2,0}
    + \sum_{cdkl} T_{ai,ck}^{2,0} \Delta^\prime_{ck,dl} T_{dl,bj}^{2,0} \nonumber \\
        &-\sqrt{\frac{\omega_\text{cav}}{2}}\sum_{ck} T_{ai,ck}^{2,0} d_{ck} T^{1,1}_{bj}
        -\sqrt{\frac{\omega_\text{cav}}{2}} T^{1,1}_{ai}
        \sum_{ck}  d_{ck} 
        T_{ck,bj}^{2,0}
        - \sqrt{\frac{\omega_\text{cav}}{2}} T^{1,1}_{ai} d_{bj} 
        - \sqrt{\frac{\omega_\text{cav}}{2}} 
        d_{ai}
        T^{1,1}_{bj}  
\end{align}
\end{widetext}
Using the definitions in Eqs.~\ref{eqn:rpa_a}, \ref{eqn:rpa_b}, and \ref{eqn:rpa_g}, we can rewrite this expression in matrix form as
\begin{align}
\label{eqn:ring_ccd_20}
    \mb{0} &= \mb{\tilde{B}}  + \mb{\tilde{A}} \mb{T}^{2,0} + \mb{T}^{2,0} \mb{\tilde{A}} + \mb{T}^{2,0}\mb{\tilde{B}}\mb{T}^{2,0} \nonumber \\
    &+ \mb{T}^{2,0} \mb{g} \left ( \mb{T}^{1,1} \right )^\dagger+ \mb{T}^{1,1} \mb{g}^\dagger \mb{T}^{2,0}  
    +  \mb{T}^{1,1} \mb{g}^\dagger + \mb{g} \left ( \mb{T}^{1,1} \right )^\dagger
\end{align}
Here, we have combined the ERI terms and DSE contributions as
\begin{align}
\mb{\tilde{A}} &= \mb{A} + \bm{\Delta} \\
    \mb{\tilde{B}} &= \mb{B} + \bm{\Delta}^\prime
\end{align}
Note that, in the absence of the cavity, the first four terms on the right-hand side of Eq.~\ref{eqn:ring_ccd_20} reduce to the residual equation for the doubles amplitudes in the standard formulation of ring-CCD.\cite{Sorensen08_231101} The matrix forms of the $\mb{T}^{1,1}$ and $T^{0,2}$ residual equations are
\begin{align}
\label{eqn:ring_ccd_11}
\mb{0} &= \mb{g} + \mb{T}^{1,1} \omega_\text{cav} + \mb{\tilde{A}} \mb{T}^{1,1} +\mb{T}^{2,0}\mb{\tilde{B}} \mb{T}^{1,1} \nonumber \\
&+ T^{0,2} \mb{g} + \mb{T}^{2,0} \mb{g}
  + \mb{T}^{1,1} \mb{g}^\dagger \mb{T}^{1,1} + T^{0,2} \mb{T}^{2,0} \mb{g}
\end{align}
and
\begin{align}
\label{eqn:ring_ccd_02}
0 = 2 \omega_\text{cav} T^{0,2} +2 \left (\mb{T}^{1,1}\right )^\dagger \mb{g}  T^{0,2} + 2\left ( \mb{T}^{1,1}\right )^\dagger\mb{g}+ \left ( \mb{T}^{1,1}\right )^\dagger \mb{\tilde{B}}\mb{T}^{1,1}
\end{align}
respectively. For expressions including explicit summations, the reader is referred to the Supporting Information. 

Given amplitudes that satisfy Eqs.~\ref{eqn:ring_ccd_20}, \ref{eqn:ring_ccd_11}, and \ref{eqn:ring_ccd_02}, the QED ring-CCD correlation energy is
\begin{align}
    E_c^\text{rCCD} &= \langle 
    0_\text{e}|\langle 0_\text{p}|{\bar{H}}|0_\text{p}\rangle | 0_\text{e}\rangle \\
    & = \frac{1}{4}\text{Tr}\left ( \mb{\tilde{B}}\mb{T}^{2,0}\right) + \mb{g}^\dagger\mb{T}^{1,1}
\end{align}
Removing the exchange integrals ($\langle pq||rs\rangle \to \langle pq|rs\rangle$) and exchange components of the DSE terms (giving $\Delta_{ai,bj} = \Delta^\prime_{ai,bj} = d_{ai}d_{bj}$) in Eqs.~\ref{eqn:ring_ccd_20}, \ref{eqn:ring_ccd_11}, and \ref{eqn:ring_ccd_02} gives us {the} ``direct'' QED-rCCD (QED-drCCD) approach. The QED-drCCD correlation energy is then given by 
\begin{align}
    E_c^\text{drCCD} = \frac{1}{2}\text{Tr}\left ( \mb{\tilde{B}}\mb{T}^{2,0}\right) + \mb{g}^\dagger\mb{T}^{1,1}
\end{align}
where $\mb{\tilde{B}}$ is also modified to exclude the exchange terms. Similarly, the direct QED-RPA (QED-dRPA) correlation energy can be obtained from Eq.~\ref{eqn:qed_rpa_energy} after solving the direct forms of Eqs.~\ref{EQN:CASIDA} and \ref{EQN:TDA}.
{
Note that the difference in the prefactor of $\text{Tr}\left ( \mb{\tilde{B}}\mb{T}^{2,0}\right)$ in the QED-rCCD and QED-drCCD correlation energies can be traced back to the QED-CCD correlation energy
\begin{align}
    E_c^\text{CCD} & {} = \frac{1}{4} \sum_{ijab} (\langle ij || ab \rangle + \Delta^\prime_{ai,bj}) T^{2,0}_{ai, bj} - \sqrt{\frac{\omega_\text{cav}}{2}}d_{ai} T^{1,1}_{ai} \nonumber\\
                   & {} = \frac{1}{2} \sum_{ijab} (\langle ij | ab \rangle + d_{ai} d_{bj} ) T^{2,0}_{ai, bj} - \sqrt{\frac{\omega_\text{cav}}{2}}d_{ai} T^{1,1}_{ai}, 
\end{align}
where we have taken advantage of the antisymmetric structure of the $T^{2,0}_{ai, bj}$ amplitude (\emph{cf.}~Refs.~\citenum{Sorensen08_231101,Toulouse11_084119,Eshuis12_1084}).

}

\subsection{The Equivalence of QED-dRPA and QED-drCCD}

{In this Section, we} derive the QED-drCCD residual equations from the QED-dRPA eigenvalue equation and show that the QED-drCCD and QED-dRPA correlation energies are equivalent. First, we regroup the electronic
 ($\mb{X}, \mb{Y}$) and photonic ($\mb{M}, \mb{N}$) degrees of freedom in the QED-dRPA problem into composite vectors for excitations 
 \begin{align}
\mathcal{U} = \begin{pmatrix} \mathbf{X} \\ \mathbf{M} \end{pmatrix}
\end{align}
and de-excitations
\begin{align}
\mathcal{V} = \begin{pmatrix} \mathbf{Y} \\ \mathbf{N} \end{pmatrix}
\end{align}
Next, we re-express Eq.~\ref{EQN:CASIDA} in a more compact form, as
\begin{align}
\label{eqn:casida_compact}
\begin{pmatrix}
\mathbb{A} & \mathbb{B} \\
-\mathbb{B} & -\mathbb{A}
\end{pmatrix}
\begin{pmatrix}
\mathcal{U} \\
\mathcal{V}
\end{pmatrix}
= 
\begin{pmatrix}
\mathcal{U} \\
\mathcal{V}
\end{pmatrix}
\bm{\Omega}
\end{align}
where
\begin{align}
\mathbb{A} = \begin{pmatrix} \mb{\tilde{A}} & \mb{g} \\ \mb{g}^\dagger & \omega_\text{cav} \end{pmatrix}
\end{align}
and
\begin{align}
\mathbb{B} = \begin{pmatrix} \mb{\tilde{B}} & \mb{\tilde{g}} \\ \mb{\tilde{g}}^\dagger & 0 \end{pmatrix}
\end{align}
As in standard RPA, the QED-RPA excitation and de-excitation vectors are related via a correlation matrix
\begin{align}
 \mathcal{T} = \mathcal{V}\mathcal{U}^{-1}
\end{align}
which has a block structure that matches that of $\mathbb{A}$ and $\mathbb{B}$, {\em i.e.} it has electron-electron, electron-photon, and photon-photon sectors, arranged as
\begin{align}
 \mathcal{T} = \begin{pmatrix}
     \mb{T}^{2,0} & \mb{T}^{1,1} \\
     (\mb{T}^{1,1})^\dagger & T^{0,2} \\
 \end{pmatrix}
\end{align}
To arrive at the QED-drCCD residual equations, we left multiply Eq.~\ref{eqn:casida_compact} by the row vector, $(\mathcal{T} \quad -\mathcal{I})$, where $\mathcal{I}$ is an identity matrix with the same dimension as $\mathcal{T}$, to obtain
\begin{align}
    \mathcal{T}\mathbb{A}\mathcal{U} + \mathbb{B}\mathcal{U} + \mathcal{T}\mathbb{B}\mathcal{V} + \mathbb{A}\mathcal{V} = \left (\mathcal{T}\mathcal{U} - \mathcal{V}\right )\bm{\Omega}
\end{align}
The right-hand side of this equation vanishes because $\mathcal{T}\mathcal{U} = \mathcal{V}$. We now multiply on the right by $\mathcal{U}^{-1}$ to obtain the Riccati equation
\begin{align}
\label{eqn:riccati}
    \mathbb{B} + \mathcal{T}\mathbb{A} + \mathbb{A}\mathcal{T} + \mathcal{T}\mathbb{B}\mathcal{T} = \mb{0}
\end{align}
similar to that in the standard RPA problem.
Equation \ref{eqn:riccati} can be partitioned into three unique components based on the block structure of $\mathbb{A}$, $\mathbb{B}$, and $\mathcal{T}$. In the electron-electron sector, we have
\begin{align}
    \mb{0} &= \mb{\tilde{B}} + \mb{\tilde{A}}\mb{T}^{2,0} +\mb{T}^{2,0}\mb{\tilde{A}} +  \mb{T}^{2,0}\mb{\tilde{B}}\mb{T}^{2,0} \nonumber \\
    & + \mb{T}^{2,0}\mb{\tilde{g}}\left ( \mb{T}^{1,1}\right)^\dagger + \mb{T}^{1,1}\mb{\tilde{g}}^\dagger\mb{T}^{2,0} + \mb{T}^{1,1} \mb{g}^\dagger + \mb{g} \left ( \mb{T}^{1,1}\right )^\dagger
\end{align}
which is equivalent to Eq.~\ref{eqn:ring_ccd_02} because $\mb{\tilde{g}} = \mb{g}$ (Eq.~\ref{eqn:rpa_g}). In the electron-photon and photon-electron sectors, we have
\begin{align}
\label{eqn:riccati_ep}
    \mb{0} &= \mb{\tilde{g}} + \mb{T}^{1,1}\omega_\text{cav} + \mb{\tilde{A}} \mb{T}^{1,1} + \mb{T}^{2,0}\mb{\tilde{B}}\mb{T}^{1,1} \nonumber \\
    &+ \mb{g} T^{0,2} + \mb{T}^{2,0}\mb{g} + \mb{T}^{2,0} \mb{\tilde{g}} T^{0,2} + \mb{T}^{1,1}\mb{\tilde{g}}^\dagger \mb{T}^{1,1}
\end{align}
and
\begin{align}
\label{eqn:riccati_pe}
\mb{0} &= \mb{\tilde{g}}^\dagger + \omega_\text{cav}\left ( \mb{T}^{1,1}\right)^\dagger + \left ( \mb{T}^{1,1}\right)^\dagger \mb{\tilde{A}} + \left ( \mb{T}^{1,1}\right)^\dagger\mb{\tilde{B}}\mb{T}^{2,0} \nonumber \\
&+T^{0,2}\mb{g}^\dagger + \mb{g}^\dagger\mb{T}^{2,0} + T^{0,2}\mb{\tilde{g}}^\dagger\mb{T}^{2,0} + \left ( \mb{T}^{1,1}\right)^\dagger \mb{\tilde{g}} \left ( \mb{T}^{1,1}\right)^\dagger
\end{align}
As expected, Eqs.~\ref{eqn:riccati_ep} and \ref{eqn:riccati_pe} are Hermitian conjugates of one another, and both are equivalent to Eq.~\ref{eqn:ring_ccd_11}. 
Lastly, in the photon-photon sector, we have
\begin{align}
    0 &= 2 \omega_\text{cav}T^{0,2}+\left ( \mb{T}^{1,1}\right)^\dagger \mb{\tilde{g}} T^{0,2} + T^{0,2} \mb{\tilde{g}}^\dagger\mb{T}^{1,1} \nonumber \\
    &+ \left ( \mb{T}^{1,1}\right)^\dagger\mb{g} + \mb{g}^\dagger\mb{T}^{1,1} + \left ( \mb{T}^{1,1}\right)^\dagger \mb{\tilde{B}} \mb{T}^{1,1}
\end{align}
which, again, is equivalent to Eq.~\ref{eqn:ring_ccd_20}. 

Having established that the QED-drCCD residual equations are derivable from the QED-dRPA eigenvalue problem, we now demonstrate the equivalence of the correlation energies from the two approaches. From Eq.~\ref{eqn:casida_compact}, we have 
\begin{align}
    \mathbb{A}\mathcal{U} + \mathbb{B}\mathcal{V} = \mathcal{U}\bm{\Omega}
\end{align}
Multiplying on the right by $\mathcal{U}^{-1}$ gives us
\begin{align}
    \mathbb{A} + \mathbb{B}\mathcal{T} = \mathcal{U}\bm{\Omega}\mathcal{U}^{-1}
\end{align}
Taking the trace of both sides and noting that the trace of $\bm{\Omega}$ is invariant to unitary transformations, leads to 
\begin{align}
    \text{Tr}\left ( \mathbb{B}\mathcal{T}\right) &= \text{Tr}\left ( \bm{\Omega}\right) - \text{Tr}\left ( \mathbb{A}\right) \nonumber \\
    & = \text{Tr}\left ( \bm{\Omega}\right) - \text{Tr}\left ( \bm{\bar{\Omega}}\right) \nonumber \\
    &= 2 E_c^\text{dRPA}
\end{align}
Expanding the trace on the left-hand side gives
\begin{align}
    \text{Tr}\left ( \mathbb{B}\mathcal{T}\right) &= \text{Tr}\left ( \mb{\tilde{B}}\mb{T}^{2,0}\right) + \text{Tr}\left ( \mb{\tilde{g}}\left [\mb{T}^{1,1}\right]^\dagger\right) + \mb{\tilde{g}}^\dagger \mb{T}^{1,1}\nonumber \\
    &= 2 E_c^\text{drCCD}
\end{align}
which completes the proof. 

\section{Numerical Validation}
\label{SEC:NUMERICAL_VALIDATION}

To numerically confirm the equivalence of cavity QED-drCCD and QED-RPA, we evaluate the ground-state correlation energy at each level of theory for a water molecule (described by the cc-pVDZ\cite{Dunning89_1007} basis set) embedded within a single-mode optical cavity. The O--H bond length is fixed at 1 \AA, the H--O--H bond angle is fixed at 104.5\degree, and the molecule is oriented such that the principal rotation axis is aligned along the $z$ direction. The molecule is coupled to a $z$-polarized single-mode optical cavity with a fundamental frequency of 0.415668 $E_\text{h}$, which is resonant with the lowest-energy excited state of the molecule with a non-zero transition moment along the $z$ direction (computed at the RPA/cc-pVDZ level, in the absence of the cavity).

The QED-RPA and QED-TDA-RWA excitation energies entering the QED-dRPA correlation energy expression were obtained by full diagonalization of the QED-RPA and QED-TDA-RWA matrices (Eqs.~\ref{EQN:CASIDA} and \ref{EQN:TDA}). The relevant code was implemented as a plugin to the \textsc{Psi4} electronic structure package.\cite{Sherrill20_184108} The excitation energies were validated against those obtained from the iterative eigensolver implementations of these methods in the \texttt{hilbert} package,\cite{hilbert} which is also a plugin to \textsc{Psi4}. The full QED-CCD equations were generated using the automated equation and code generation tool, \texttt{p}$^\dagger$\texttt{q}.\cite{DePrince21_e1954709,DePrince25_6679} The ring simplifications of these equations were made by hand (as described the Supporting Information) and implemented in an in-house Python code, with QED-HF orbitals generated using the QED-HF implementation in \texttt{hilbert} and integrals taken from \textsc{Psi4}.

{Table 1 provides total QED-HF energies and QED-drCCD / QED-RPA correlation energies at different coupling strengths ($\lambda$, given in a.u.). One can see that, over the range of experimentally accessible coupling strengths\cite{Baumberg16_127} ({\em i.e.}, $\lambda \le 0.05$ a.u.), the mean-field energy increases by roughly 5.0 m$E_\text{h}$, due to the presence of the DSE term. The correlation energies become more negative by $\approx 1.2$ m$E_\text{h}$ over the same range, partially canceling the increase in energy at the mean-field level. This trend is consistent with our previous observations that mean-field cavity QED methods over-estimate the energetic impact of the cavity; methods that incorporate explicit electron-photon interactions predict milder effects.\cite{DePrince21_094112,DePrince23_5264,DePrince24_064109}}

Table \ref{tab:ecorr} {also} provides differences in correlation energies obtained from QED-drCCD and QED-dRPA. As expected, the correlation energies for the two methods are essentially identical, agreeing to within $1\times10^{-12}$ $E_\text{h}$ for all coupling strengths. Table \ref{tab:ecorr} also quantifies the influence of the photon-photon channel by providing the difference between ground-state correlation energies obtained from a QED-drCCD model that ignores $T^{0,2}$ in the cluster operator and the full QED-drCCD approach. As we can see, photon-photon correlation effects are on the order of 1 $\mu E_\text{h}$ or less for physically realizable coupling strengths ($\lambda \le 0.05$ a.u.). The $T^{0,2}$ channel becomes more energetically important at larger coupling strengths, even exceeding 1 m$E_\text{h}$ for $\lambda \ge 0.4$ a.u. We can also quantify the relative importance of  the $T^{0,2}$ channel by calculating what percentage of the change in correlation energy due to the presence of the cavity ({\em i.e.}, relative to $\lambda = 0.0$ a.u.) originates from the this term. Again, we see that $T^{0,2}$ is fairly unimportant at feasible coupling strengths, never exceeding $0.1\%$ for $\lambda \le 0.05$ a.u. The relative importance of this channel grows to a few percent for the largest coupling strengths we consider.

\begin{table}[!htpb]
    \centering
    \caption{{Total QED-HF energies ($E^\text{HF}$, $E_\text{h}$),  QED-dRPA correlation energies ($E_\text{h}$), t}he difference in the correlation energies obtained from QED-drCCD and QED-dRPA {($E_\text{h}$),} and the change in the correlation energy when ignoring the $T^{0,2}$ channel ($E_\text{h}$). The numbers in parentheses represent the percentage of the  change in correlation energy relative to $\lambda = 0.0$ a.u. originating from the  $T^{0,2}$ channel. }
    \label{tab:ecorr}
    \begin{tabular}{ccccc}
    \hline
    \hline
    $\lambda$ & $E^\text{HF}$ & $E_c^\text{dRPA}$ & $E_c^\text{drCCD}-E_c^\text{dRPA}$ & $T^{0,2}$ channel \\
    \hline
    0.00	& $-$76.021418	& $-$0.233459 & $-6.2 \times$ 10$^{-13}$ &	0.0                    (0.00\%)\\
    0.01	& $-$76.021216	& $-$0.233508 & $-4.6 \times$ 10$^{-14}$ &	1.8$\times$ 10$^{-09}$ (0.00\%)\\
    0.02	& $-$76.020608	& $-$0.233654 & $+8.3 \times$ 10$^{-14}$ &	2.9$\times$ 10$^{-08}$ (0.01\%)\\
    0.03	& $-$76.019594	& $-$0.233898 & $-1.1 \times$ 10$^{-13}$ &	1.4$\times$ 10$^{-07}$ (0.03\%)\\
    0.04	& $-$76.018177	& $-$0.234238 & $-2.2 \times$ 10$^{-13}$ &	4.6$\times$ 10$^{-07}$ (0.06\%)\\
    0.05	& $-$76.016355	& $-$0.234676 & $-2.7 \times$ 10$^{-13}$ &	1.1$\times$ 10$^{-06}$ (0.09\%)\\
    0.10	& $-$76.001237	& $-$0.238305 & $+1.2 \times$ 10$^{-13}$ &	1.7$\times$ 10$^{-05}$ (0.35\%)\\
    0.20	& $-$75.941747	& $-$0.252591 & $-4.3 \times$ 10$^{-13}$ &	2.2$\times$ 10$^{-04}$ (1.17\%)\\
    0.30	& $-$75.845567	& $-$0.275896 & $+8.6 \times$ 10$^{-13}$ &	8.8$\times$ 10$^{-04}$ (2.07\%)\\
    0.40	& $-$75.715887	& $-$0.307993 & $-2.6 \times$ 10$^{-14}$ &	2.1$\times$ 10$^{-03}$ (2.78\%)\\
    0.50	& $-$75.555889	& $-$0.348717 & $-8.7 \times$ 10$^{-13}$ &	3.7$\times$ 10$^{-03}$ (3.22\%) \\
    \hline
    \end{tabular}
\end{table}

\section{Conclusions}

\label{SEC:CONCLUSIONS}

We have developed an analytic proof of the equivalence of the ground-state correlation energy from QED-RPA and QED-ring-CCD{. Given what is known about the electron-only version of this problem and the similar structures of the CC and QED-CC wave functions, this result may have been anticipated, but the present proof nonetheless confirms} QED-RPA's position in the QED-CC heirarchy and {establishes} it as a tool for describing the ground states of strongly-coupled light--matter systems. A key feature of the QED-ring-CCD model used here is its inclusion of double photon creation terms, which are often overlooked in QED-CC studies. The presence of this channel even at the QED-RPA level suggests that the associated effects should be included in other approximate QED-CC models to ensure that this limit is recovered correctly. Moreover, the QED-RPA and QED-ring-CCD methods analyzed in this work incorporate explicit electron-photon correlation effects that are essential for accurate descriptions of cavity-induced changes to electronic structure; such changes are often overestimated by mean-field approaches that ignore electron-photon correlation.\cite{DePrince21_094112,DePrince23_5264,DePrince24_064109} As such, by leveraging the known favorable physical properties of RPA in standard electronic structure theory,\cite{Toulouse05_012510,Rubio06_073201,Kurth00_16430,Burke05_094116} as well as its computational efficiency,\cite{Furche08_114105,Furche12_084105,Scheffler12_7447} QED-RPA is well-positioned to enable accurate, large-scale simulations of cavity-modified properties in molecules and materials. As an efficient tool for describing strong light--matter interactions, the QED-RPA will serve as a useful complement to existing density-functional-theory-based approaches that either ignore electron-photon correlation effects  or approximate them via electron-photon correlation functionals.\cite{Rubio15_093001,Rubio17_113036,Rubio18_992,Rubio21_e2110464118,Tokatly23_235424,Rubio24_052823,Flick25_073002}

\vspace{0.5cm}
{\bf Supporting Information} Programmable expressions for the full QED-CCD residual equations and the ring simplifications thereof. Correlation energies computed using QED-dRPA and QED-drCCD.

\begin{acknowledgments}
SHY acknowledges support from the FSU Quantum Initiative.
\end{acknowledgments}

\vspace{0.5cm}

\noindent {\bf DATA AVAILABILITY}\\

The data that support the findings of this study are available from the corresponding author upon reasonable request.

\bibliography{main.bib}

\end{document}